# Autonomous and Collaborative Smart Home Security System (ACSHSS)

Hassan Jalil Hadi , Khaleeq Un Nisa, Sheetal Harris
Bahria University, Islamabad, Pakistan
Email: hassanjalilhadi1142@gmail.com

*Abstract*— **In the modern era of the internet, billions of IoT devices are connected to share valuable information across the networks. According to McKinsey Global Institute, there are twenty-seven IoT devices per second becoming part of the internet and it is expected that by 2025, the number will reach up to 64 billion IoT devices globally (research by Gartner). Such an enormous number of devices brings the concept of house automation and monitoring. House automation enables an individual to take timely actions in case of any emergency. Today's world is the victim of cybercrimes especially; IoT networks are a paradise for attackers. By considering all these challenges, in this research paper, the aim is to design, deploy and implement a secure house automation system. Firstly, the proposed solution provides remotely accessible integrated IoT resources for the safety and security of the building. By using Sha ort Messaging System (SMS), the age is sent to the user by the Global System for Mobile (GSM) system. An SMS alert is sent to the user in case any sensor detects an abnormality in their operation. Secondly, an authentication mechanism is deployed to enable only authorized users to access resources. Thirdly, in case of a malicious approach in accessing IoT resources, a timely alert should be received by the owner. A Network Intrusion Detection System (NIDS) is deployed to detect and real-time inform in case of any suspicious activity while accessing the Internet of Things network.**

## I. INTRODUCTION

Today, we live in the fourth industrial revolution which is changing perfect human life from domestic level to commercial activities. Internet of Things (IoT), robotics and artificial intelligence (AI) are prime technologies that play a significant role to ease human livings. Internet of Things (IoT) is one of the tremendous and major innovations of the internet. It is firstly used by Kevin Ashton in 1999 at the Massachusetts Institute of Technology (MIT). It always remains a human desire to utilize innovative technologies to ease their lives. Internet of Things transformed the concept of traditional network devices to nontraditional dimensions. IoT devices are used to carry important data over the internet [1]. An individual can get a lot of information about others based on internet-connected devices. Besides all these advantages, there are some disadvantages and back draws.

Cybersecurity is one of the major concerns and a challenge of advanced internet-connected technologies. Smart internet of things can be used to access and monitor home security. As the internet of things has limited resources, which is one of the major motivational factors for the hacker to intrude on the network and steal the information. There is always a presence of risk about intrusion or hazard condition which requires an authentic entry over the wire or physically. The proposed solution is an autonomous and collaborative smart security system for IoT based home systems. The proposed system deployed and installed multiple sensors for the purpose of physical detection. GSM module is deployed to send an alert timely and to make the client vigilant to initiate remedial actions.

Authentication is the process the process or action of verifying the identity of a user or process, the response is generated as per output of that authentication. This technology provides access control for systems by checking to see if a user's credentials match the credentials in a database of authorized users or a data authentication server [2]. The ever-increasing size of the Internet of Things (IoT) put serious challenge over its secure authorization and authentication mechanism which cope through open authorization (OAuth) framework [3]. It is suggested to combine authentication and authorization mechanism to restrict the client. Additionally, in the proposed solution in this paper, there is a mechanism of authentication and authorization to access resources to authentic clients only. Network intrusion detection systems (NIDS) are set up at the network entry point to examine traffic on all devices in the network [4]. It performs an observation of passing traffic and matches it to the collection of known attacks. Once an attack is identified or abnormal behavior is observed, the alert can be sent to the administrator.

This research paper comprises of multiple sections. Section II comprises of related work. Section III illustrates the proposed solution to a secure home automation system. Section IV describes the deployment and implementation of the proposed solution. Section V wraps up the result analysis. Section VI conclusion the research paper.

## II. RELATED WORK

The significance of cybersecurity is rapidly increasing all around the world. Whereas peoples are more concerned about malware, worm, viruses or spams and adware now. While public, private and government organizations dealing in information security are aware of the situation, Public awareness is the prime factor in information security. When it comes to home or building automation, the public awareness is very low. Statistical data of suspicious and illegal network activities is a key factor in influencing public opinion and the importance of research and analysis. Many systems are designed for security purposes such as,

In 2015, Chandana et.al [5] proposed a smart surveillance system using things to speak and Raspberry pi for home and office security. The model uses hardware mechanisms such as Raspberry Pi (model B), Gyro sensor and Raspberry pi camera. This system will monitor when motion detected, the Raspberry Pi will control the Raspberry Pi camera to take a picture and sent out an email alert. The proposed system work in standalone mode without the requirement of PC once programmed.

Andrea et.al [6] provided definitions and visions for the smart office of tomorrow, which refers to the office setting. The main contribution of the paper is a concept for the management and use of smart office service devices, which defines how smart devices can be integrated into a smart office environment and how the devices can be managed and controlled through a network calendar and a voice service like Amazon Alexa. With this, we illustrate how assistance systems like Amazon Alexa are gradually opening their closed platforms and how this can be used to develop a solution for smart IoT security.

In 2014, Sanjana et.al [7] proposed to design and implementation of a smart surveillance monitoring system using Raspberry Pi and PIR sensors for mobile devices. It increases the usage of mobile technology to provide essential security to our homes and for other control applications. The proposed home security system captures information and transmits it via a 3G Dongle to a smartphone using a web application. Raspberry pi operates and controls motion detectors and video cameras for remote sensing and surveillance, streams live video and records it for future playback. It can also find the number of persons located with the help of the infrared sensor.

When it comes to security of IoT devices many techniques have been proposed. For example, a honeypot is a system or service in a network without any real application or use. The main goal of a honeypot is to detect an attack on the network. There have been numerous efforts of detected malware in IoT Devices. Javid et al, [8] proposed an IDS system that uses a whitelist to prevent IoT devices from connecting to malicious addresses and avoid communications with botnets Command and Control or private data leaks. However, tests with real botnet attacks were not carried out, and the IDS depends on the maintenance of the third-party systems.

Likewise, the author N-BaIoT [9], proposed a NIDS that used deep auto encoders to detect botnets in IoT devices and uses network traffic of IoT devices to build a model of legitimate behavior and detect any anomaly. They achieved great results in detecting the attacks, with a low false positives rate. Despite the great results, the use of deep auto encoders can be computationally costly and demands large amounts of data.

Furthermore, El-Hajj et al in 2019 [10], Introduced an examination of numerous mechanism for authentication in the Internet of Things (IoT) like constrained climate, it dependent on various measures like breaks down and compares about existing validation conventions by introducing their advantages and disadvantages. The recommended research paper depends on a survey. Processing power which is obliged to the IoT climate.

In addition, Cirani et al in 2014 [12], proposed to save and make sure about administrations for IoT gadgets dependent on OAuth verification and approval structure. The proposed framework is best because of lower processing capacity, execute modified admittance strategies, adaptable and lightweight. The proposed arrangement utilized a solitary confirmation factor that can cause a solitary purpose of failure. As it is obvious from above discussion that all above strategies are for general security yet not for smart home automation.

There is a lack of NIDS and authentication based smart and secure home security. Existing techniques have the limitation of a single security factor that is vulnerable to security attacks. Hence, there is a great need to design and implement a multi-factor secure network for home IoT devices. This solution will provide safety, security, stability, and sustainability to IoT devices. The proposed solution provides authentication and NIDS detection to the IoT network.

Section II contains related work. Proposed solution for a protected home automation system is described in Section III. Section IV depicts the deployment and implementation of the proposed system. In Section results are analyzed. Conclusion of the research work is presented in Section VI.

### III. PROPOSED SYSTEM

This section briefly elaborates the proposed solution for autonomous and collaborative smart home security system. The system consists of four sensors including, fire sensor, two motion sensors, and gas sensor. It also has IP camera. All sensors are centrally connected and controlled by the system controller known as Raspberry PI. If any of the sensor receives any detection, it sends it to the system controller. After receiving detection from sensor, the system controller generates an alert and sends an SMS/Call notification to house owner through GSM. The IP camera captures the picture of the hurdles and sends it to the Web application. Camera also sends the livestream to the webserver. This home security is remotely accessible by the owner. Due to remote access facility additional security measures were needed. For this, a NIDS is deployed between the system controller and the public network. Any malicious attempt is timely detected by NIDS to initiate preventive measures. The IoT resources like camera streaming, pictures, and sensor inputs are available on the webserver so it was required that only authorized person should be able to access the web application. For this, an authentication mechanism is deployed to verify the authorized client. When the client wants to access resources, firstly, the client provides login credentials. After verification of the client's credentials, the key is received via email for the next authentication step. Client enter received a key, if the key is verified then the client can successfully access the resources.

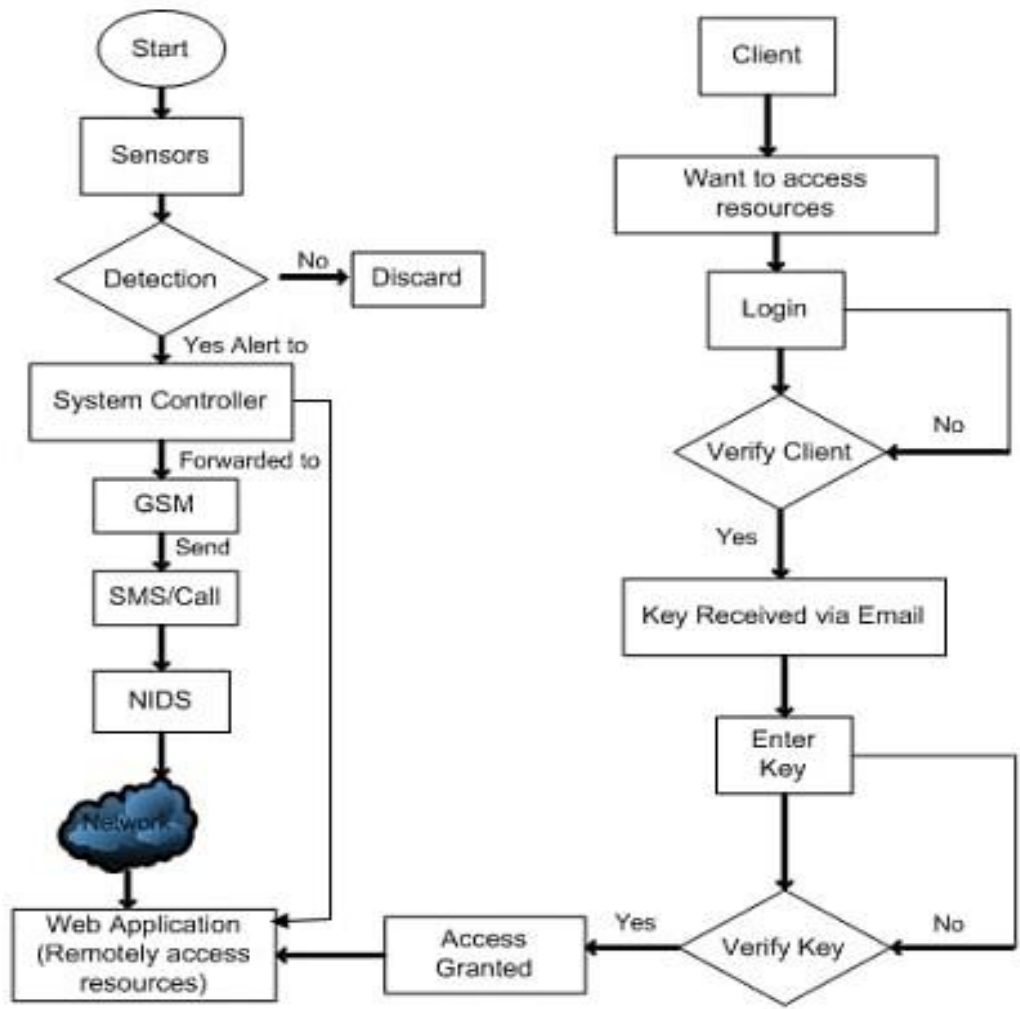

Fig. 1. Workflow ACSHSS (Autonomous and Collaborative Smart Home Security System)

## IV. DEPLOYMENT

Section IV explains the deployment of the proposed solution. For the deployment of smart home security system NIDS and authentication mechanism is deployed successfully. The proposed solution is comprised of IoT sensors, system controller, GSM module, NIDS, and an authentication server. Client requests to access IoT devices remotely. Firstly, authentication is required for confirmation of a valid client. An authentication mechanism is subdivided into two parts. In the earliest part of authentication, the client should provide its credentials which are verified. In the subsequent part of authentication, the client received key via email which is verified. In case of verification of both steps, the client successfully accesses the web application interface but to access home IoT devices a NIDS is deployed between the system controller and public network. NIDS is deployed for the sole purpose of securing IoT devices installed in-home infrastructure. NIDS detects malicious attempts performed by the attacker. The client received an alert via SMS/Call to perform preventive measures as shown in fig 2.

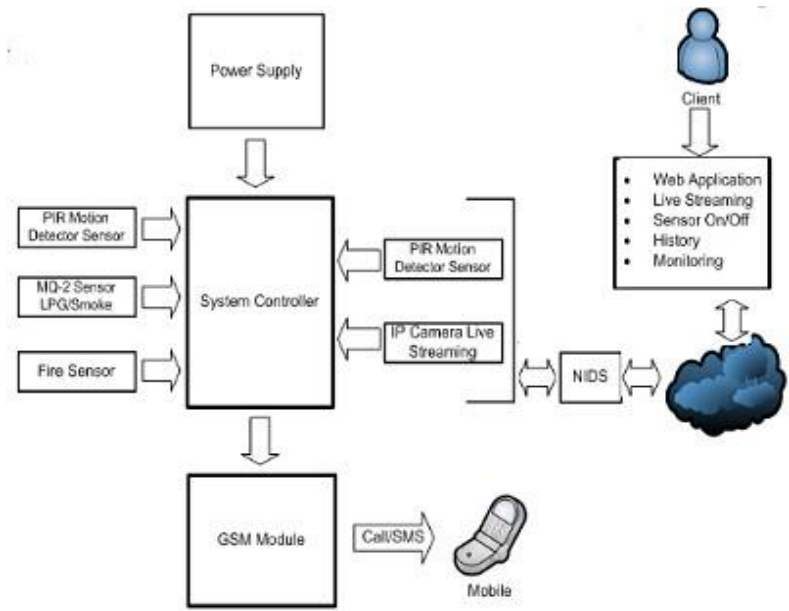

Fig. 2. System level Deployment

## V. IMPLEMETATION RESULTS

Proposed system consists of following steps.

- Embedding Sensors with System Controlled
- Lightweight Network Intrusion Detection System
- Home security Web Application
- Authentication Mechanism

### A. Embedding Sensors with System Controller

All sensors were connected and controlled by the system controller known as Raspberry PI. When a sensor received a detection, it forwarded detection to the system controller. After receiving detection from sensor, the system controller generated an alert and sent an SMS notification to house owner through GSM. The IP camera captured the picture and livestreamed the hurdle and send it to the Web application.

### B. Lightweight Network Intrusion Detection System

For security purpose a NIDs was designed which is aimed at developing a prototype that could provide a layer of security against intruders and could also be managed by owner of the system. In case of any malicious activity by system controller is notified through an email to homeowner. The implementation results for this module are given below.

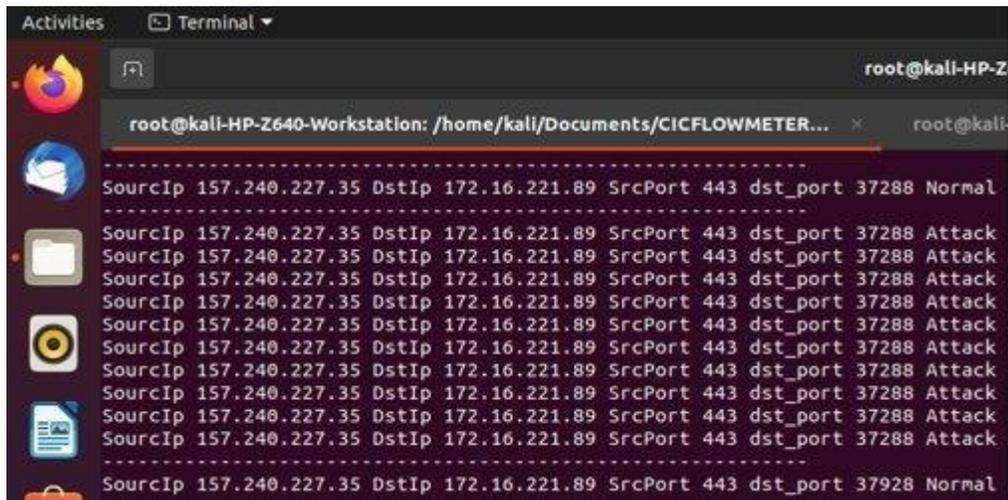

Fig. 3. Attack Detection

### C. Web Application for Home Security System

A web application is developed to save and view the history of the sensors. It shows when a sensor was turned on or off and when an intrusion was detected. Pictures captured and lives streaming by IP camera are saved in database which is linked to web application. The main purpose of this module is to facilitate the owner with remote access. So that he can monitor the detection/behavior after receiving the notification when he is outside the home. The results are given below,

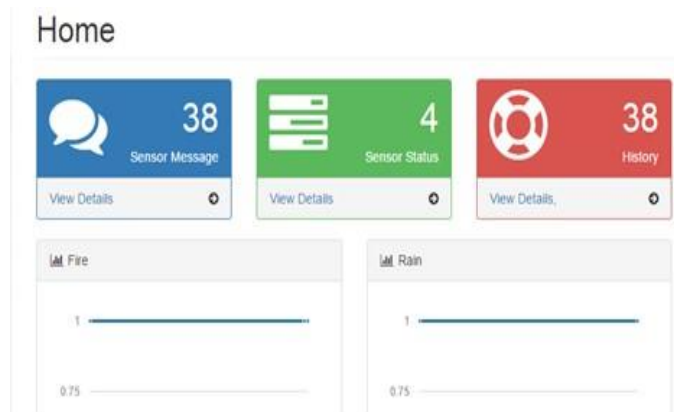

Fig. 4. Web Application for Home Security System

### D. Authentication Mechanism

To secure IoT resources, an authentication method is used to prevent a compromise. The attacker tries to compromise resources, but the authentication mechanism restricts him. Two factors authentication is implemented. The first is token verification and second is key distribution. Firstly, the client provides login credentials. After verification of the client's credentials, the key is received via email for the next authentication step. Client enter received key, if the key is verified then the client can successfully access the resources. If key is not verified then it is required to enter again a valid key. The output is shown below.

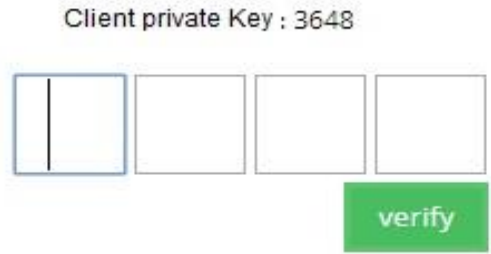

Fig. 5. Key verification process

## VI. PERFOMANCE ANALYSIS

After the successful performance of the authentication the proposed system was tested in order to analyze its performance. This performance analysis helps to evaluate the efficiency of the proposed system. The system was tested in two rounds. At first the performance of authentication module was tested. For this, an attacker model was deployed using Mirai Botnet. Numerous attacks were performed to check the efficiency of authentication mechanism. After testing it was concluded that sometimes login credential was compromised but key distribution mechanism was not compromised. The performance is shown in the graph below.

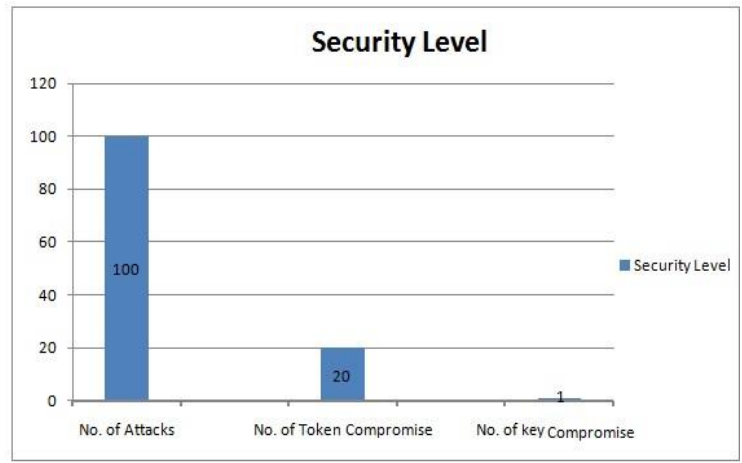

Fig. 6. Authentication mechanism performance

In second phase, lightweight network intrusion detection system was tested. The Bot-IoT dataset which has normal IoT network Traffic and numerous attackers was used for evaluation of proposed the solution system. The main reason for selecting this data type its show realistic IoT echo environment. The Dataset contains DDoS, DoS, OS and services scan and keylogging attacks. To identify network level pattern for diverse kind of traffic that device create and used these patterns to detect suspicious behavior. The main Characteristic and description of attack are presented in tables while statistically attack as well as normal behavior shown in the figure below,

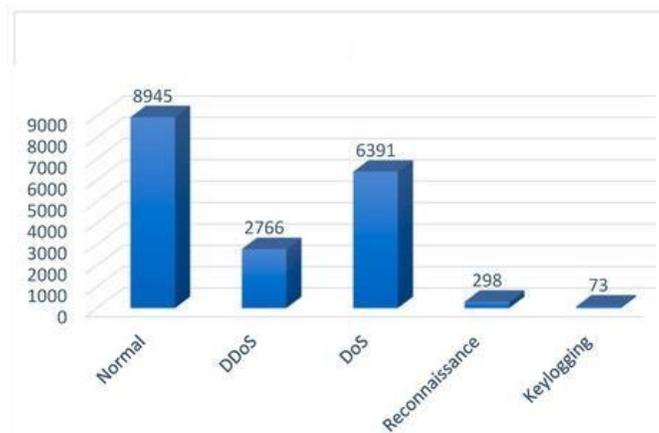

Fig. 7. NIDS performance

## VII. CONCLUSION

IoT devices transport sensitive and secure information over the internet. Due to immense dependence upon IoT Devices, it has encouraged the attacker to compromise it. DDoS reply and impersonation attacks are launched over IoT. The proposed Autonomous and Collaborative Smart Home Security System (ACSHSS) tries to restrict attacks by enabling authentication and NIDS. In the results, it has been proven that our method has made the usage of IoT devices more secure and user friendly. Authentication and NIDS are working in harmony to provide a secure and reliable working mechanism. In the future, this research can be enhanced and tested for zero-day attacks on IoT devices and for other new attacks.